\documentstyle[12pt]{article}
\newcommand{\be}{\begin{equation}}
\newcommand{\ee}{\end{equation}}
\newcommand{\ba}{\begin{eqnarray}}
\newcommand{\ea}{\end{eqnarray}}
\newcommand{\no}{\nonumber}

\def\beq{\begin{equation}}
\def\eeq{\end{equation}}
\def\bea{\begin{eqnarray}}
\def\eea{\end{eqnarray}}
\def\bq{\begin{quote}}
\def\eq{\end{quote}}

\def\gappeq{\mathrel{\rlap {\raise.5ex\hbox{$>$}}
{\lower.5ex\hbox{$\sim$}}}}

\def\lappeq{\mathrel{\rlap{\raise.5ex\hbox{$<$}}
{\lower.5ex\hbox{$\sim$}}}}

\def\Toprel#1\over#2{\mathrel{\mathop{#2}\limits^{#1}}}

\begin{document}
\pagestyle{empty}
\begin{flushright}
{CERN-TH/2002-016}\\
hep-th/0201250\\
\end{flushright}
\vspace*{5mm}
\begin{center}
{\large{\bf Schwinger terms in gravitation in two dimensions\\ as a consequence of the gravitational anomaly\footnote{This work was supported by the Austria-Czech Republic Scientific Callaboration, Project No. 2001-11.}}} \\
\vspace*{1cm}

{\bf Emmanuel Kohlprath}\footnote{e-mail address: Emmanuel.Kohlprath@cern.ch} \\

\vspace{0.3cm}

Theoretical Physics Division, CERN \\
CH - 1211 Geneva 23 \\
\vspace*{2cm}
{\bf ABSTRACT} \\ \end{center}
\vspace*{5mm}
\noindent
We compute the Schwinger term in the gravitational constraints in two dimensions, starting from the path integral in Hamiltonian form and the Einstein anomaly.

\vspace*{5cm}
\noindent

\noindent

\vspace*{0.5cm}

\vfill\eject
%\pagestyle{empty}
%\clearpage\mbox{}\clearpage

\setcounter{page}{1}
\pagestyle{plain}

\section{Introduction}
A Yang--Mills theory with a non-abelian anomaly (gauge anomaly) leads to Schwinger terms (central charges) in the constraint algebra (the Gauss law operators) as well as in the algebra of currents (see e.g. \cite{Jackiw85},\cite{Zumino1985}). Theories of gravitation and matter that have a gravitational anomaly (Einstein or Lorentz) also lead to Schwinger terms in the constraints and currents (energy--momentum tensors). As we will see later, in the tensor theory of Einstein in two dimensions, the gravitational constraints reduce to the energy--momentum tensor and therefore the two Schwinger terms are equivalent. This case has been considered in \cite{Tomiya86} -- \cite{BertlmannKohlprath3}. The Schwinger terms in the constraints of a scalar--tensor theory are discussed in \cite{CangemiJackiw94} -- \cite{Jackiw97}.

In \cite{AlekseevMadaichikFaddeevShatashvili87}, Faddeev et al. found the following method to compute the Schwinger term in the algebra of the Gauss law operators in a Yang--Mills theory: starting with the path integral in Hamiltonian form, they make a gauge transformation and include the non-abelian anomaly. From the Ward identity in second order in the gauge parameter, one can then extract the Schwinger term by acting with a suitable operator. Our goal is to generalize this to gravitation.
\vspace*{-30pt}

\section{Gravitation as a constrained Hamiltonian system}

We start with Einstein's theory of gravitation and a massless chiral fermion, which are described by the action (we have eighter $P_+\psi=0$ or $P_-\psi=0$ with $P_\pm=\frac{1}{2}(1\pm\gamma_5)$):
\be
S=\int\!\!dx\,e\left[R+\frac{i}{2}e^{a\mu}\bar\psi\gamma_a\stackrel{\leftrightarrow}{\nabla}_\mu\psi\right].\label{action}
\ee
This action can also be written as a constrained Hamiltonian system (see e.g. \cite{DeserIsham76} -- \cite{Teitelboim83a}) (in $d$-dimensions $i$ runs from $1$ to $d-1$):
\be
S=\int\!\!dx\,\left[\pi_a^{\ i}\dot e^a_{\ i}+i\ \ \ g^{\hspace*{-16pt}(3)\hspace*{6pt}1/2}\ \bar\psi\Gamma^\bot\dot\psi-N\tilde{\mathcal{H}}_\bot-N^i\tilde{\mathcal{H}}_i-\frac{1}{2}\omega_0^{\ ab}J_{ab}\right],
\ee
where $\ \ \ g^{\hspace*{-16pt}(3)\hspace*{6pt}ij}\ $ is the induced metric and $\pi_a^{\ i}$ is the canonical momentum to $e^a_{\ i}$ defined by $p^{ij}=\frac{1}{4}\left(\pi_a^{\ i}\ \ \ e^{\hspace*{-14pt}(3)\hspace*{4pt}aj}\ +\pi_a^{\ j}\ \ \ e^{\hspace*{-14pt}(3)\hspace*{4pt}ai}\right)$, where the canonical momentum to $g_{ij}$ is expressed in terms of the extrinsic curvature $K_{ij}$
\be
p^{ij}=\ \ \ g^{\hspace*{-16pt}(3)\hspace*{6pt}1/2}\ (K^{ij}-K \ \ \ g^{\hspace*{-16pt}(3)\hspace*{6pt}ij}\ )\label{conjugatemomentumingravitation}.
\ee
The Lagrangian multipliers are the lapse $N$, the shift $N_i$, $N^i=\ \ \ g^{\hspace*{-16pt}(3)\hspace*{6pt}ij}\ N_j$, and the $0$-component of the spin connection $\omega_0^{\ ab}$. The constraints are
\ba
{\tilde{\mathcal{H}}_{\bot}}&=&{\mathcal{H}}_{\bot}+\partial_k J^{\bot k}-i\ \ \ g^{\hspace*{-16pt}(3)\hspace*{6pt}1/2}\ \bar\psi\Gamma^iD_i\psi\\
\tilde{\mathcal{H}}_i&=&{\mathcal{H}}_i+\frac{1}{2}g_{ij}\partial_kJ^{kj}+K_{ik}J^{\bot k}+i\Bigl[\ \ \ g^{\hspace*{-16pt}(3)\hspace*{6pt}1/2}\ \bar\psi\Gamma^\bot D_i\psi-\frac{1}{4}\partial_k\Bigl(\ \ \ g^{\hspace*{-16pt}(3)\hspace*{6pt}1/2}\ \bar\psi[\Gamma_i,\Gamma^k]\Gamma^\bot\psi\Bigr)\Bigr]\\
J_{ab}&=&\pi_a^{\ i}e_{bi}-\pi_b^{\ i}e_{ai}-\frac{i}{4}\ \ \ g^{\hspace*{-16pt}(3)\hspace*{6pt}1/2}\ \bar\psi\Bigl(\Gamma^\bot\gamma_a\Gamma^i-\Gamma^i\gamma_a\Gamma^\bot\Bigr)e_{bi}+\frac{i}{4}\ \ \ g^{\hspace*{-16pt}(3)\hspace*{6pt}1/2}\ \bar\psi\Bigl(\Gamma^\bot\gamma_b\Gamma^i-\Gamma^i\gamma_b\Gamma^\bot\Bigr)e_{ai}\no\\&&\\
\Gamma^i&=&\ \ \ g^{\hspace*{-16pt}(3)\hspace*{6pt}ij}e^a_{\ j}\ \gamma_a=\ \ \ e^{\hspace*{-14pt}(3)\hspace*{6pt}ai}\gamma_a,\qquad \Gamma^\bot=-n^a\gamma_a.
\ea
Remember that the gravitational Hamiltonian is vanishing.
\section{The gravitational path integral}
The general coordinate transformations are the gauge transformations of gravitation. To define a well behaved path integral, we choose the gauge fixing $e^a_{\ 0}=\delta^a_{\ 0}$ and introduce the coordinate ghosts using the Faddeev--Popov method. Under infinitesimal active coordinate transformations (Einstein transformations) the vielbein transforms as
\be
\delta^c_\xi e^a_{\ \mu}(x)=e^{'a}_{\mu}(x)-e^a_\mu(x)=\xi^\nu\partial_\nu e^a_{\ \mu}(x)+e^a_{\ \nu}(x)\partial_\mu\xi^\nu.
\ee
This leads to the ghost action
\ba
S_{GH}&=&\int\!\!dx\,dy\,e(x)\bar v^\nu(x)e_{a\nu}(x)\left.\frac{\delta^c_\xi(e^a_{\ 0}-\delta^a_{\ 0})}{\delta \xi_\mu(y)}\right|_{\xi=0}v_\mu(y)\no\\&=&\int\!\!dx\,e(x)\bar v_\nu(x)\left(e_a^{\ \nu}(x)\partial_\mu e^a_{\ 0}(x)v^\mu(x)+\partial_0 v^\nu(x)\right).
\ea
We find the path integral in Hamiltonian form for gravitation and a chiral fermion
\ba
Z&=&\frac{1}{N}\int\!\!d\pi_a^{\ i}\,de^a_{\ i}\,de^a_{\ 0}\,d\bar\psi\, d\psi\, d\bar v_\alpha dv^\alpha \delta(e^a_{\ 0}-\delta^a_{\ 0}) \no\\&&\times\exp \Bigl\{i\int\!\!dx\,\Bigl[\pi_a^{\ i}\dot e^a_{\ i}+i\ \ \ g^{\hspace*{-16pt}(3)\hspace*{6pt}1/2}\ \bar\psi\Gamma^\bot\dot\psi-N^\mu\tilde{\mathcal{H}}_\mu-\frac{1}{2}\omega_0^{\ ab}J_{ab}\no\\&&+e\,\bar v_\nu\left(e_a^{\ \nu}\partial_\mu e^a_{\ 0}v^\mu+\partial_0 v^\nu\right)\Bigr]\Bigr\}.\label{Hamiltonianpathintegralforgravitationandfermions}
\ea
The inclusion of powers of $e$ as weights in the fermionic measure, as it is explained in \cite{Fujikawa83} and \cite{FujikawaYasuda84}, would be no problem but we will not need it.

\section{Schwinger terms in gravitation in two dimensions}

In two dimensions the Einstein--Hilbert action is proportional to the Euler number of the two-dimensional manifold (see e.g. \cite{Deser96}). We choose a manifold where the Euler number vanishes. So in two dimensions there are no dynamical degrees of freedom for gravity in Einstein's theory. From (\ref{conjugatemomentumingravitation}) we see that the conjugate momenta $\pi_a^{\ 1}$ identically vanish. Using $e^a_{\ 1}e^b_{\ 1}[\gamma_a,\gamma_b]=0$ and $\gamma_b\gamma_0\gamma_a-\gamma_a\gamma_0\gamma_b=0$ the constraints reduce to
\ba
\tilde{\mathcal{H}}_{\bot}&=&-i\ \ \ g^{\hspace*{-16pt}(1)\hspace*{6pt}1/2}\ \bar\psi\Gamma^1\partial_1\psi\\
\tilde{\mathcal{H}}_1&=&i\ \ \ g^{\hspace*{-16pt}(1)\hspace*{6pt}1/2}\ \bar\psi\Gamma^\bot \partial_1\psi\\
J_{ab}&=&0.
\ea
We have
\ba
\Gamma^1&=&\ \ e^{\hspace*{-14pt}(1)\hspace{4pt}a1}\gamma_a=(g_{11})^{-1}e^a_{\ 1}\gamma_a=e^{a1}\gamma_a-\frac{N^1}{N}n^a\gamma_a\\
\Gamma^\bot&=&N^\bot e^{a0}\gamma_a=-n^a\gamma_a\\
e&=&|g|^{1/2}=\ \ \ g^{\hspace*{-16pt}(1)\hspace*{6pt}1/2}\ N,
\ea
so that
\be
N^\mu{\mathcal{H}}_\mu=-ie\,\bar\psi e^{a1}\gamma_a\partial_1\psi.\label{gravitationalconstraintsintwodimensions}
\ee
The path integral (\ref{Hamiltonianpathintegralforgravitationandfermions}) reduces to
\ba
Z&=&\frac{1}{N}\int\!\!d\pi_a^{\ 1}de^a_{\ 1}\,de^a_{\ 0}\,d\bar\psi\, d\psi\, d\bar v_\alpha dv^\alpha \delta(e^a_{\ 0}-\delta^a_{\ 0}) \no\\&&\times\exp \Bigl\{i\int\!\!dx\,\Bigl[\pi_a^{\ 1}\dot e^a_{\ 1}+ie\,\bar\psi e^{a0}\gamma_a\partial_0\psi+ie\,\bar\psi e^{a1}\gamma_a\partial_1\psi\no\\&&+e\,\bar v_\nu\left(e_a^{\ \nu}\partial_\mu e^a_{\ 0}v^\mu+\partial_0 v^\nu\right)\Bigr]\Bigr\}.
\ea
We relabel all fields, $e^a_{\ \mu}(x)\rightarrow e'^a_{\ \mu}(x)$, as for the other fields. We interpret this as an active coordinate transformation and use the invariance of the classical action and the bosonic measure under coordinate transformations. The fermionic measure gives us the Einstein anomaly $G[\Lambda,\Gamma]$, and we are left with
\ba
Z&=&\frac{1}{N}\int\!\!d\pi_a^{\ 1}de^a_{\ 1}\,de'^a_{\ 0}\,d\bar\psi\, d\psi\, d\bar v_\alpha dv^\alpha \delta(e'^a_{\ 0}-\delta^a_{\ 0}) \no\\&&\times\exp \Bigl\{i\int\!\!dx\,\Bigl[\pi_a^{\ 1}\dot e^a_{\ 1}+ie\,\bar\psi e^{a0}\gamma_a\partial_0\psi+ie\,\bar\psi e^{a1}\gamma_a\partial_1\psi\no\\&&+e\,\bar v_\nu\left(e_a^{\ \nu}\partial_\mu e^a_{\ 0}v^\mu+\partial_0 v^\nu\right)\Bigr]\Bigr\}G[\Lambda,\Gamma]\label{Hamiltonianpathintegralintwodimensions}
\ea
with the explicit expression (see e.g. \cite{Bertlmann})
\be
G[\Lambda,\Gamma]=\exp\Bigl\{\pm\frac{i}{96\pi}\Bigl[\frac{1}{3}\int\limits_{\Gamma^+}\textrm{tr}(d\Lambda \Lambda^{-1})^3+\int\limits_{M_2}\textrm{tr} (d\Lambda\Lambda^{-1}\Gamma)\Bigr]\Bigr\},\label{Einsteinanomaly}
\ee
where $\Gamma^a_{\ b}$ is the Christoffel connexion 1-form, $\Lambda^a_{\ b}$ is the ``gauge'' element and $\partial\Gamma^+=M_2$. For a general coordinate transformation $x'=x'(x)$ we define the ``gauge'' parameter $\xi$ by
\ba
x'^\alpha&=&x^\alpha-\xi^\alpha(x)\\
(\Lambda^{-1})^\alpha_{\ \beta}&=&\frac{\partial x'^\alpha}{\partial x^\beta}=\delta^\alpha_{\ \beta}-\partial_\beta\xi^\alpha(x)
\ea
so that, in second order in $\xi$, we have
\ba
x^\alpha&=&x'^\alpha+\xi^\alpha(x')+\xi^\lambda\partial_\lambda\xi^\alpha(x')+O(\xi^3)\\
\Lambda^\alpha_{\ \beta}&=&\frac{\partial x^\alpha}{\partial x'^\beta}=\delta^\alpha_{\ \beta}+\partial_\beta\xi^\alpha(x)+\partial_\beta\xi^\lambda\partial_\lambda\xi^\alpha(x)+O(\xi^3).
\ea
The zweibein transforms under passive coordinate transformations as
\be
e'^a_{\ \mu}(x')=\Lambda^\nu_{\ \mu}e^a_{\ \nu}(x)=e^a_{\ \mu}(x)+\partial_\mu\xi^\nu e^a_{\ \nu}(x)+\partial_\mu\xi^\lambda\partial_\lambda\xi^\nu e^a_{\ \nu}(x)+O(\xi^3)
\ee
and under active coordinate tranformations (Einstein transformations) as
\ba
e'^a_{\ \mu}(x)&=&e^a_{\ \mu}(x)+\partial_\mu\xi^\nu e^a_{\ \nu}(x)+\xi^\alpha\partial_\alpha e^a_{\ \mu}(x)+\partial_\mu\xi^\lambda\partial_\lambda\xi^\nu e^a_{\ \nu}(x)\no\\&&+\xi^\alpha\partial_\alpha\partial_\mu\xi^\nu e^a_{\ \nu}(x)+\xi^\alpha\partial_\mu\xi^\nu\partial_\alpha e^a_{\ \nu}(x)+\xi^\alpha\partial_\alpha\xi^\beta\partial_\beta e^a_{\ \mu}(x)\no\\&&+\frac{1}{2}\xi^\alpha\xi^\beta\partial_\alpha\partial_\beta e^a_{\ \mu}(x)+O(\xi^3).
\ea
The gauge fixing $e'^a_{\ 0}(x)=\delta^a_{\ 0}$ leads to a second order differential equation for $e^a_{\ 0}$. If we are in two dimensions and choose $\xi^1=0$, then we find
\be
\delta^a_{\ 0}=\Bigl[1+\partial_0\xi^0+\xi^0\partial_0+\partial_0\xi^0\partial_0\xi^0+\xi^0\partial_0\partial_0\xi^0+2\xi^0\partial_0\xi^0\partial_0+\frac{1}{2}\xi^0\xi^0\partial_0\partial_0\Bigr]e^a_{\ 0}.
\ee
Its solution is
\be
e^a_{\ 0}=(1-\partial_0\xi^0)\delta^a_{\ 0}+O(\xi^3).\label{gaugefixedvielbein}
\ee
Using (\ref{gaugefixedvielbein}) we express everything up to second order in $\xi$, in terms of $e^0_{\ 1}$ and $e^1_{\ 1}$. For the Einstein anomaly (\ref{Einsteinanomaly}) we find
\ba
G[\xi,\Gamma]&=&\exp\Bigl\{\pm\frac{i}{48\pi}\int\!\!dx\,\Bigl[\partial_0\partial_0\xi^0\partial_1\partial_0\xi^0(e^0_{\ 1})^2(e^1_{\ 1})^{-2}+\partial_1\partial_0\xi^0\partial_1\partial_1\xi^0(e^1_{\ 1})^{-2}\no\\&&-2\partial_1\partial_0\xi^0\partial_1\partial_0\xi^0e^0_{\ 1}(e^1_{\ 1})^{-2}+\Bigr[\Bigl(\partial_0\partial_0\xi^0+2\partial_0\partial_0\xi^0\partial_0\xi^0\Bigr)(e^0_{\ 1})^2(e^1_{\ 1})^{-2}\no\\&&-2\Bigl(\partial_1\partial_0\xi^0+\partial_1\partial_0\xi^0\partial_0\xi^0\Bigr)e^0_{\ 1}(e^1_{\ 1})^{-2}+\partial_1\partial_1\xi^0(e^1_{\ 1})^{-2}\Bigr]\partial_0e^0_{\ 1}\no\\&&+\Bigr[-\Bigr(\partial_0\partial_0\xi^0+2\partial_0\partial_0\xi^0\partial_0\xi^0\Bigr)e^0_{\ 1}(e^1_{\ 1})^{-1}\no\\&&+\Bigl(\partial_1\partial_0\xi^0+\partial_1\partial_0\xi^0\partial_0\xi^0\Bigr)(e^1_{\ 1})^{-1}\Bigr]\partial_0e^1_{\ 1}\Bigr]\Bigr\}+O(\xi^3)\label{Einsteinanomaly2},
\ea
where we use the convention $\varepsilon^{01}=1$. As can be seen from (\ref{Hamiltonianpathintegralintwodimensions}), the momentum $\pi_a^{\ 1}$ is no longer conjugate to $e^a_{\ 1}$, since there are terms in (\ref{Einsteinanomaly2}) proportional to $\partial_0e^0_{\ 1}$ and $\partial_0e^1_{\ 1}$. Therefore we make a shift in $\pi_a^{\ 1}$ to absorb these terms. The functional determinant is simply $1$ and its effect is to kill all terms proportional to $\partial_0e^0_{\ 1}$ and $\partial_0e^1_{\ 1}$ in (\ref{Einsteinanomaly2}). Using
\ba
{\mathcal{L}}_{GH}&=&e\,\bar v_\nu\left(e_a^{\ \nu}\partial_\mu e^a_{\ 0}v^\mu+\partial_0 v^\nu\right)=e\,\bar v_\nu\partial_0 v^\nu+e^1_{\ 1}\partial_\mu e^0_{\ 0}\bar v_0 v^\mu\\
e\,e^{a0}\gamma_a&=&-(e^1_{\ 1}\gamma_0+e^0_{\ 1}\gamma_1), \qquad e\,e^{a1}\gamma_a=e\,(e^1_{\ 1})^{-1}\gamma_1
\ea
we find the generating functional in second order in the gauge parameter
\ba
Z&=&\frac{1}{N}\int\!\!d\pi_a^{\ 1}de^a_{\ 1}\,d\bar\psi\,d\psi\,d\bar v_\alpha dv^\alpha\exp \Bigl\{i\int\!\!dx\,\Bigl[\pi_a^{\ 1}\dot e^a_{\ 1}-i\bar\psi(e^1_{\ 1}\gamma_0+e^0_{\ 1}\gamma_1)\partial_0\psi\no\\&&+\Bigl(1-\partial_0\xi^0\Bigr)i\bar\psi\gamma_1\partial_1\psi+\Bigl(1-\partial_0\xi^0\Bigr)e^1_{\ 1}\bar v_\nu\partial_0 v^\nu-\partial_\mu\partial_0\xi^0e^1_{\ 1}\bar v_0 v^\mu\no\\&&\pm\frac{1}{48\pi}\Bigl[\partial_0\partial_0\xi^0\partial_1\partial_0\xi^0(e^0_{\ 1})^2(e^1_{\ 1})^{-2}-2\partial_1\partial_0\xi^0\partial_1\partial_0\xi^0e^0_{\ 1}(e^1_{\ 1})^{-2}\no\\&&+\partial_1\partial_0\xi^0\partial_1\partial_1\xi^0(e^1_{\ 1})^{-2}\Bigl]\Bigl]\Bigl\}+O(\xi^3).
\ea
In $O((\xi^0)^2)$ we find the following Ward identity
\ba
0&=&\left\langle\pm\frac{i}{48\pi}\int\!\!dx\,\Bigl[\partial_0\partial_0\xi^0\partial_1\partial_0\xi^0(e^0_{\ 1})^2(e^1_{\ 1})^{-2}-2\partial_1\partial_0\xi^0\partial_1\partial_0\xi^0e^0_{\ 1}(e^1_{\ 1})^{-2}\right.+\partial_1\partial_0\xi^0\partial_1\partial_1\xi^0(e^1_{\ 1})^{-2}\Bigl]\no\\&&\left.+\frac{1}{2}\Bigl[\int\!\!dx\Bigl[\partial_0\xi^0\bar\psi\gamma_1\partial_1\psi-i\partial_0\xi^0 e^1_{\ 1}v_\nu\partial_0 v^\nu-i\partial_\mu\partial_0\xi^0e^1_{\ 1}\bar v_0 v^\mu\Bigr]\Bigr]^{2}\right\rangle.\label{2ndorderwardidentityforgravity}
\ea
Acting with $\frac{\delta^2}{\delta\xi^0(x)\delta\xi^0(y)}$ on (\ref{2ndorderwardidentityforgravity}) we arrive at
\ba
0&=&\left\langle\pm\frac{i}{48\pi}\Bigl[\Bigl(2\partial_1^y\partial_0^y\partial_0^y\partial_0^y\delta(x-y)+3\partial_1^y\partial_0^y\partial_0^y\delta(x-y)\partial_0^y+\partial_0^y\partial_0^y\partial_0^y\delta(x-y)\partial_1^y\right.\no\\&&+\partial_0^y\partial_0^y\delta(x-y)\partial_1^y\partial_0^y+\partial_1^y\partial_0^y\delta(x-y)\partial_0^y\partial_0^y\Bigr)\Bigl((e^0_{\ 1})^2(e^1_{\ 1})^{-2}\Bigr)\no\\&&-4\Bigl(\partial_1^y\partial_1^y\partial_0^y\partial_0^y\delta(x-y)+\partial_1^y\partial_1^y\partial_0^y\delta(x-y)\partial_0^y+\partial_1^y\partial_0^y\partial_0^y\delta(x-y)\partial_1^y\no\\&&+\partial_1^y\partial_0^y\delta(x-y)\partial_1^y\partial_0^y\Bigr)\Bigl(e^0_{\ 1}(e^1_{\ 1})^{-2}\Bigr)+\Bigl(2\partial_1^y\partial_1^y\partial_1^y\partial_0^y\delta(x-y)\no\\&&+\partial_1^y\partial_1^y\partial_1^y\delta(x-y)\partial_0^y+3\partial_1^y\partial_1^y\partial_0^y\delta(x-y)\partial_1^y+\partial_1^y\partial_1^y\delta(x-y)\partial_1^y\partial_0^y\no\\&&\left.+\partial_1^y\partial_0^y\delta(x-y)\partial_1^y\partial_1^y\Bigr)(e^1_{\ 1})^{-2}\Bigr]\right\rangle\no\\&&+\partial_0^x\partial_0^y\left\langle\Bigl(-\bar\psi\gamma_1\partial_1\psi(x)+ie^1_{\ 1}\bar v_\nu\partial_0 v^\nu(x)+i\partial_\mu^x(e^1_{\ 1}\bar v_0 v^\mu(x))\Bigr)\right.\no\\&&\left.\times\Bigl(-\bar\psi\gamma_1\partial_1\psi(y)+ie^1_{\ 1}\bar v_\nu\partial_0 v^\nu(y)+i\partial_\mu^y(e^1_{\ 1}\bar v_0 v^\mu(y))\Bigr)\right\rangle.\label{2ndorderwardidentityforgravity2}
\ea
The last two lines give the commutator times $\partial_0^y\delta(x^0-y^0)$:
\be
\partial_0^y\delta(x-y)\Bigl[\bar\psi\gamma_1\partial_1\psi(x),\bar\psi\gamma_1\partial_1\psi(y)\Bigr],
\ee
plus terms that are proportional to $\delta(x^0-y^0)$ or regular as $y^0\rightarrow x^0$. Next we apply
\be
\lim\limits_{(p_0-q_0)\rightarrow\infty}\frac{p_0-q_0}{p_0q_0^3}\int\!\!dx^0\,dy^0\,e^{ip_0x^0+iq_0y^0}
\ee
on (\ref{2ndorderwardidentityforgravity2}) to project onto terms proportional to $\partial_0^y\partial_0^y\partial_0^y\delta(x-y)$ and we find
\be
0=\pm\frac{i}{48\pi}\langle\Bigl(2\partial_1^y\delta(x^1-y^1)+\delta(x^1-y^1)\partial_1^y\Bigr)\Bigl((e^0_{\ 1})^2(e^1_{\ 1})^{-2}\Bigr)\rangle.
\ee
Using this, we apply
\be
\lim\limits_{(p_0-q_0)\rightarrow\infty}\frac{p_0-q_0}{p_0q_0^2}\int\!\!dx^0\,dy^0\,e^{ip_0x^0+iq_0y^0}
\ee
on (\ref{2ndorderwardidentityforgravity2}) to project onto terms proportional to $\partial_0^y\partial_0^y\delta(x-y)$. We obtain
\ba
0&=&\left\langle\pm\frac{i}{48\pi}\Bigr[\partial_1^y\delta(x^1-y^1)\partial_0^y\Bigl((e^0_{\ 1})^2(e^1_{\ 1})^{-2}\Bigr)-4\Bigl(\partial_1^y\partial_1^y\delta(x^1-y^1)\right.\no\\&&\left.+\partial_1^y\delta(x^1-y^1)\partial_1^y\Bigr)\Bigl(e^0_{\ 1}(e^1_{\ 1})^{-2}\Bigr)\Bigr]\right\rangle.
\ea
Using this and finally applying
\be
\lim\limits_{(p_0-q_0)\rightarrow\infty}\frac{p_0-q_0}{p_0q_0}\int\!\!dx^0\,dy^0\,e^{ip_0x^0+iq_0y^0}
\ee
on (\ref{2ndorderwardidentityforgravity2}) to project onto terms proportional to $\partial_0^y\delta(x-y)$, we arrive at
\ba
\left\langle\Bigl[-i\bar\psi\gamma_1\partial_1\psi(x),-i\bar\psi\gamma_1\partial_1\psi(y)\Bigr]\right\rangle=\pm\frac{i}{48\pi}\left\langle\Bigl[2\partial_1^y\partial_1^y\partial_1^y\delta(x^1-y^1)\right.\no\\ \left.+3\partial_1^y\partial_1^y\delta(x^1-y^1)\partial_1^y+\partial_1^y\delta(x^1-y^1)\partial_1^y\partial_1^y\Bigr](e^1_{\ 1})^{-2}\right\rangle.
\ea
From the fermionic part of the action (\ref{action}) we find the energy--momentum tensor
\be
T_{\mu\nu}=-\frac{i}{4}\Bigl[\bar\psi\gamma_\mu\stackrel{\leftrightarrow}{\nabla}_\nu\psi+\bar\psi\gamma_\nu\stackrel{\leftrightarrow}{\nabla}_\mu\psi\Bigr]+g_{\mu\nu}\frac{i}{2}\bar\psi\gamma^\lambda\stackrel{\leftrightarrow}{\nabla}_\lambda\psi.
\ee
For flat space, we have $e^0_{\ 1}=0,e^1_{\ 1}=1$, and we find the Schwinger term in the energy--momentum tensor
\be
\langle[T_{00}(x),T_{00}(y)]_{ET}\rangle=\pm\frac{i}{24\pi}\partial_1^y\partial_1^y\partial_1^y\delta(x^1-y^1).
\ee
This indeed agrees with \cite{Tomiya86} -- \cite{BertlmannKohlprath3}.

\section{Conclusion}

From (\ref{gravitationalconstraintsintwodimensions}) we see that we found an elegant way to compute the Schwinger term in the gravitational constraints in two dimensions, which emphasizes its relation to the gravitational anomaly. The gravitational anomaly contributes in $4k+2=2,6,10,\dots$ dimensions and we expect that this method could be generalized to these higher dimensions. However, the calculation will become more complicated there.

\end{document}